%% file: Article_IEEE_IS2_2025_IWMM_DAW_in_MR.tex
\documentclass[conference]{IEEEtran}
\IEEEoverridecommandlockouts

\usepackage{todonotes}
\usepackage{cite}
\usepackage{amsmath,amssymb,amsfonts}
\usepackage{algorithmic}
\usepackage{graphicx}
\usepackage{textcomp}
\usepackage{xcolor}
\def\BibTeX{{\rm B\kern-.05em{\sc i\kern-.025em b}\kern-.08em
    T\kern-.1667em\lower.7ex\hbox{E}\kern-.125emX}}
\begin{document}

\title{MR-DAW: Towards Collaborative Digital Audio Workstations in Mixed Reality\\
\thanks{This work has been supported by the MUSMET project funded by the EIC Pathfinder Open scheme of the European Commission (grant agreement n. 101184379).}
}

\author{
\IEEEauthorblockN{Torin Hopkins}
\IEEEauthorblockA{\textit{SolJAMM Research}\\
Denver, USA \\
0000-0001-7359-2906}
\and
\IEEEauthorblockN{Shih-Yu Ma}
\IEEEauthorblockA{\textit{University of Colorado Boulder} \\
Boulder, USA \\
0000-0001-6686-0079}
\and
\IEEEauthorblockN{Suibi Che-Chuan Weng}
\IEEEauthorblockA{\textit{ATLAS Insitute} \\
\textit{University of Colorado, Boulder}\\
Boulder, USA \\
chwe1250@colorado.edu}
\and
\IEEEauthorblockN{Ming-Yuan Pai}
\IEEEauthorblockA{\textit{Hsinchu County Liou-Jia Senior High School} \\
\textit{National Taiwan Normal University}\\
Taipei, Taiwan \\
maggie93198@gmail.com}
\and
\IEEEauthorblockN{Ellen Yi-Luen Do}
\IEEEauthorblockA{\textit{ATLAS Institute} \\
\textit{University of Colorado, Boulder}\\
Boulder, USA \\
ellen.do@colorado.edu}
\and
\IEEEauthorblockN{Luca Turchet}
\IEEEauthorblockA{\textit{Dep. of Information Engineering } \\
\textit{and Computer Science}\\
\textit{University of Trento}\\
Trento, Italy \\
luca.turchet@unitn.it}
}

\maketitle

\begin{figure*}[!t]
    \centering
    \includegraphics[width=\textwidth]{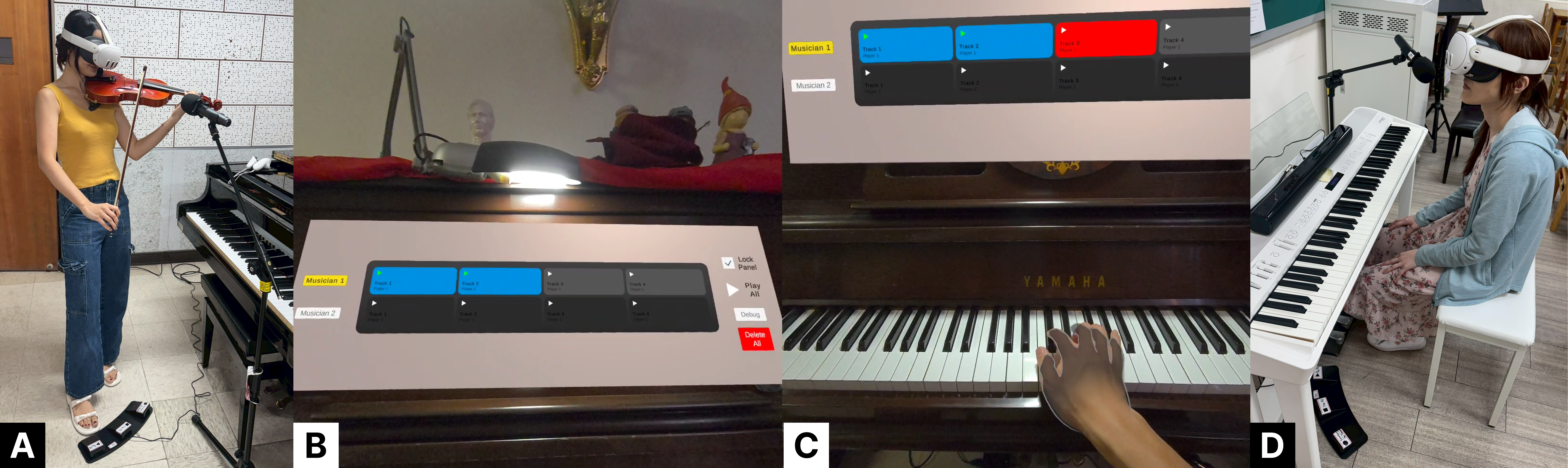}
    \caption{Overview of the MRDAW mixed reality music collaboration system. (A) User n. 1 playing violin with a foot pedal for recording. (B) Mixed reality view from user showing the spatialized UI panel with track controls overlaid on the physical environment. (C) MR view from user demonstrating track recording feedback (red) and the spatial UI integrated into the physical space. (D) User n. 2 playing piano with a foot pedal for recording. Both users wear Meta Quest 3 headsets.}
    \label{fig:system_overview}
\end{figure*}

\IEEEpeerreviewmaketitle

\begin{abstract}
Digital Audio Workstations (DAWs) are central to modern music production but often encumber the musician's workflow, tethering them to a desk and hindering natural interaction with their instrument. Furthermore, effective remote collaboration remains a significant challenge, with existing solutions hampered by network latency and asynchronous file sharing. This paper investigates the potential of Mixed Reality (MR) to overcome these barriers, creating an intuitive environment for real-time, remote musical collaboration. We employ qualitative and speculative design techniques to better understand: 1) how players currently use DAWs, and 2) to imagine a speculative future of collaborative MR-DAWs. To facilitate this discussion, we developed and evaluated the usability of a design probe, MR-DAW. An MR system enabling multiple, geographically dispersed users to control a single, shared DAW instance while moving freely in their local spaces. Our networked system enables each remote musician to use a physical foot pedal for collaborative looping, merging a familiar, hands-free interaction with a shared virtual session. Based on interviews and system evaluations with 20 musicians, we analyze current practices, report on the user experience with our MR system, and speculate on the future of musical collaboration in MR. Our results highlight the affordances of MR for unencumbered musical interaction and provide a speculative outlook on the future of remote collaborative DAWs in the Musical Metaverse.
\end{abstract}


\begin{IEEEkeywords}
Musical Metaverse, Musical XR, Digital Audio Workstation, encumbered interactions
\end{IEEEkeywords}

\section{Introduction}
\input{01_Intro}

\section{System description}
\input{02_system}

\section{Interaction}
\input{03_interaction}

%
%

\section{Evaluation}
\input{04_Evaluation}

\section{Discussion}
\input{05_Discussion}

\section{Conclusions}

\input{06_Conclusion}


\section*{Acknowledgment}

We acknowledge the support of the MUSMET project funded by the EIC Pathfinder Open scheme of the European Commission (grant agreement n. 101184379). Views and opinions expressed are however those of the author(s) only and do not necessarily reflect those of the European Union or the European Innovation Council. Neither the European Union nor the European Innovation Council can be held responsible for them. 


\bibliographystyle{IEEEtran}
\bibliography{musical_XR.bib,Turchet.bib,NMP.bib,IoMusT.bib,article_specific.bib, other.bib}

\end{document}

%% file: 01_Intro.tex
Recording music has long been accomplished with the teamwork of a musician and an engineer. However, advances in computation and music technology have ushered in a new era of music recording, empowering musicians to record their music easily using only a personal computer. This method of recording music has been widely adopted in the industry for not only creating new music, but also sharing musical ideas with others and preparing material for live performance. Musicians now incorporate the computer into their musical workflow for all of these purposes and more, including: musical practice, reading music, and playing music remotely, requiring access to digital material while simultaneously making music or holding a musical instrument.

While engaged in such activities, musicians often find themselves positioned at a desk holding an instrument and working on a computer simultaneously. This requires the use of a mouse and keyboard, or other music devices for controlling digital audio workstations (DAWs)--computer programs that facilitate recording and many other musical practices \cite{marrington2017composing}. For many musicians, this positioning is a direct hindrance to their workflow, creating obstruction and distance between them and their instrument \cite{avila2019encumbered}. 

The inherent inconvenience of this interaction paradigm, coupled with the computational demand of professional grade DAWs makes collaborative composition a difficult process. Those collaborating on musical compositions are often relegated to file sharing asynchronously or working around a desktop with very little spatial flexibility. Though networked collaborative DAWs do exist \cite{abletonlink}, they are either constrained by local area networks \cite{abletonlink}, or in the case of remote DAWs \cite{barate2022collaborative}, are either latency-ridden, require turn taking, or have very limited real-time file sharing capabilities \cite{stickland2021design}. 


Extended Reality, which encompasses Virtual Reality (VR) and Mixed Reality (MR), is rapidly emerging as a new space where conducting music activities, including collaborative composition \cite{turchet2021MusicalXR,hopkins2024xrjam, ARDC_hopkins}. 
Various DAWs have been proposed in both VR and MR \cite{turchet2023freesoundvr,cassorla2020augmented,men2018lemo}. However, to our best knowledge, existing systems using traditional instruments have focused on single-user interactions, as they were not conceived for usage in multi-user settings. This falls in the remits of the Musical Metaverse research field \cite{turchet2023musical}.

By using the MR medium it is possible to provide collaborating musicians with the visual experience of interacting with a shared DAW while also enabling them to move about in a free and unencumbered manner. To complement this freedom, it is possible to provide musicians with external devices to control aspects of the recording process. These include foot pedals, leading to an interaction mechanism familiar to many musicians, such as the common pedal board of guitar players. These newfound interactions for music in shared MR environments enable musicians to move about their space and organize interfaces in a customized manner, untethering the musician from the desktop work environment that has been demonstrated to interfere with the creative process \cite{avila2019encumbered}.

MR systems are uniquely positioned to overcome the limitation of collaboration on a DAW by providing multiple players with independent interfaces that control a single computer-based DAW. This differs from the current paradigm which entails updating two or more systems simultaneously with data-intensive file exchange or asynchronous file sharing. The simultaneous control of a single DAW using MR enables multiple players to send control messages and near-real-time audio streams--two functions already available to musicians within certain geographical and network limitations thanks to the usage of Networked Music Performance systems  \cite{rottondi2016overview}. This allows us to overcome traditional obstacles in remote collaborative composition, paving the way for the emergence of the Musical Metaverse. 

In this paper, we explore the affordances of MR for DAW-based collaborative composition as compared to computer-based workflows. Firstly, we conducted a set of interviews in which 20 participants described their current experiences using technology as an aid to practice, perform, and record music. Based on this understanding, we then introduced participants to a DAW in MR we engineered to exemplify the common practice and performance technique of looping--layering music on the fly using a foot pedal. After utilizing the system, participants were asked to reflect on the use of MR in collaborative composition, play, and practice, and to speculate on the possibilities for future use of DAWs in MR.    

Our study contributes new insight into the nature of collaboration in MR-based music environments. Through innovative collaborative looping and speculative design we contribute the following:

\begin{enumerate}
    \item We present \textit{MR-DAW}, a MR-based collaborative DAW design probe which enables players to create and record new music that interfaces directly with a computer-based DAW; 
    \item We described a collaborative networked loop pedal; 
    \item We report the results of the thematic analysis of the current workflow of 20 musicians;
    \item We discuss the results of thematic analysis pertaining to the use of current technology and a speculative future for collaborative DAWs in the Musical Metaverse. 
\end{enumerate}


%% file: 02_system.tex
\begin{figure*}
    \centering
    \includegraphics[width=\linewidth]{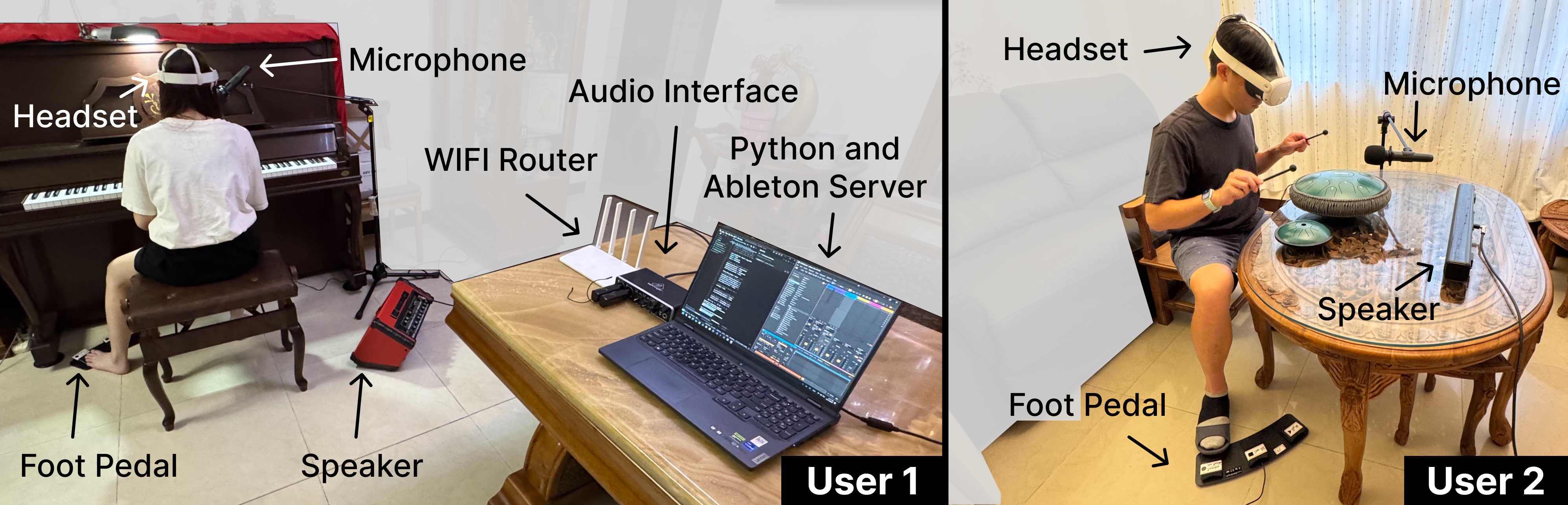}
    \caption{Two distinct user setups for interacting with a Mixed Reality Digital Audio Workstation (MR DAW). ``User 1" (left) interacts with a virtual piano using a headset, while a microphone captures audio. The setup includes a laptop displaying the DAW software (Ableton Live and Python), an audio interface, a WIFI router, a speaker, and a foot pedal. ``User 2" (right) engages with virtual drums using drumsticks and a headset. Their setup includes a microphone, a speaker, and a foot pedal on the floor. Both configurations demonstrate different methods of musical creation and interaction within a mixed reality environment.}
    \label{fig:system_setup}
\end{figure*}

Our design probe, MR-DAW, is conceived for two users, although it can be easily extended to a higher number of users (Fig.~\ref{fig:system_setup}).
 The system comprises three core components: a MR user interface, a foot pedal, an audio interface, and a Python-based Ableton Live controller. This architecture ensures seamless integration of virtual interactions, real-time audio, and professional DAW capabilities. In the following, the three components are detailed.



\subsection{Mixed Reality User Interface}
Developed with Unity 3D and deployed on Meta Quest 3 headsets, the MR panel is the primary visual and interactive hub (see Figure~\ref{fig:system_overview}C, D). It includes:
\begin{itemize}
    \item \textbf{Eight Audio Tracks:} Visually represent loop tracks, providing real-time feedback on recording (red), playback (green), and selection (blue);
    \item \textbf{Session Control Buttons:} Spatial UI toggles mirroring physical foot pedal functions for global playback and individual track toggling;
    \item \textbf{Debug Window:} A discrete panel for system diagnostics (network status, audio levels, errors).
\end{itemize}
This spatialized UI enables natural, embodied interaction, freeing musicians from traditional desk setups.

\subsection{Audio Interface and Communication}
A Behringer U-Phoria UMC204HD audio interface manages audio I/O, capturing microphone input and delivering synchronized playback to each user. Each station includes a stand-mounted microphone and speaker, positioned in separate physical space for remote collaboration simulation. 

\subsection{Ableton Live Controller}
Ableton Live 12 orchestrates the musical session. A custom Python controller manages user inputs and automates Live functions, bridging MR interactions and Live. The central laptop, connected to the Behringer audio interface, runs Live. Open Sound Control (OSC) facilitates low-latency network communication between headsets and a central laptop (running Ableton Live) \cite{jones2023abletonosc}. The Python script communicates with Live (via API/OSC) and foot pedals. Each user operates dedicated foot pedals for tactile control over:
\begin{itemize}
    \item \textbf{Track Recording:} Initiating and stopping individual loop recordings;
    \item \textbf{Playback:} Global play/stop commands.
\end{itemize}
This setup ensures precise synchronization and control, allowing musicians to focus on performance.

%% file: 03_interaction.tex
\begin{figure*}
    \centering
    \includegraphics[width=\linewidth]{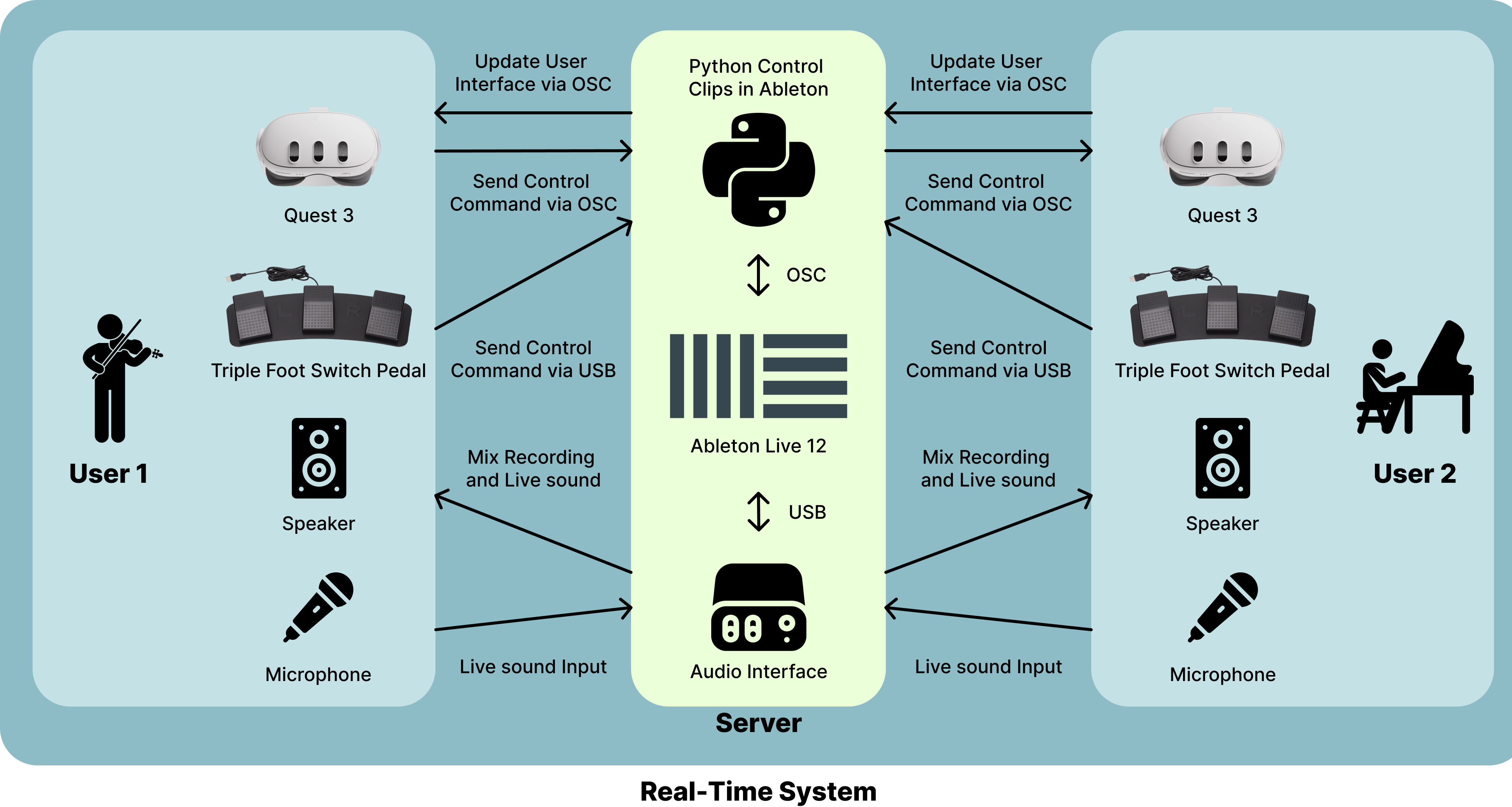}
    \caption{The system architecture of the real-time, collaborative mixed reality music environment. The system is centered around a server running Ableton Live 12 and a Python control script, which communicate via Open Sound Control (OSC). Each user is equipped with a Quest 3 headset, a triple foot switch pedal, a microphone, and a speaker. Control commands from the headsets (via OSC) and foot pedals (via USB) are sent to the Python script on the server. The script processes these inputs to manipulate Ableton Live and sends visual feedback to the users' headset interfaces via OSC. Simultaneously, live audio from each user's microphone is routed to a central audio interface. Ableton Live processes and mixes this live sound with digital audio, and the final mix is sent back to each user's speaker.}
\label{fig:system_diagram}
\end{figure*}

MR-DAW facilitates collaborative music creation through three primary categories of interaction within its loop-based environment: loop track recording, track playback control, and real-time communication. These interactions are designed to be intuitive and support a seamless co-creation workflow for musicians in a mixed reality setting, utilizing a combination of gesture and extensible controllers (Fig.~\ref{fig:system_diagram}).

\subsection{Loop Track Recording}
Each user within the MR-DAW system is allocated the ability to record up to four distinct audio tracks. The recording process is initiated and terminated through a dedicated foot pedal, offering a tactile and embodied control method that allows musicians to maintain focus on their instruments or performance. While a track is actively being recorded, its corresponding visual element on the spatial MR UI panel provides immediate feedback by turning red (see Figure~\ref{fig:system_overview}D), indicating its active status. Upon the second press of the foot pedal, the system automatically ceases recording, quantizes and loops the newly captured audio, and subsequently activates the next available empty track, preparing it for a subsequent recording.

A critical design decision for collaborative synchronization is that the \textbf{loop duration is determined by the first track recorded by user n. 1}. This establishes a foundational tempo and loop length for the entire collaborative session. Consequently, all subsequent tracks recorded by either user are automatically trimmed or extended to precisely match this initial loop duration, ensuring perfect rhythmic synchronization across all layered parts. This feature is crucial for maintaining a cohesive musical structure in a real-time collaborative jamming scenario. Once both users have filled their four allocated tracks (eight tracks total), the system automatically resets to the first track. This mechanism allows users to continuously overwrite existing loops or layer new recordings, enabling iterative composition and dynamic musical exploration within the fixed loop structure.

\subsection{Track Playback Control}
Users have comprehensive control over playback, utilizing both the physical foot pedal and the spatial mixed reality UI panel. The foot pedal is equipped with two dedicated buttons: one for initiating playback of all active tracks and another for stopping all tracks (see Figure~\ref{fig:system_overview}A, B). The UI panel mirrors these functionalities with readily accessible toggle buttons, providing an alternative visual control method. Furthermore, the spatial UI offers individual toggle buttons for each track, allowing musicians to selectively enable or disable specific loops. This granular control is essential for focused listening during collaborative review sessions, enabling users to isolate parts, experiment with arrangements, and refine their mix in real-time.

\subsection{Real-Time Communication}
To foster authentic and spontaneous musical collaboration, MR-DAW incorporates a robust real-time communication system. The system continuously monitors each user's microphone input and transmits the live audio feed directly to the other user's speaker. This creates an open audio channel that enables natural verbal communication and allows musicians to hear each other's live instrumental performances, even when located in physically separate rooms. This setup is vital for spontaneous musical interaction, supporting co-creation with physical instruments such as a guitar, keyboard, or even vocals, closely mimicking the experience of jamming in the same physical space.

%% file: 04_Evaluation.tex
\subsection{Evaluation Protocol}
Our evaluation is based on semi-structured interviews of the 10 pairs of participants, and thematic analysis of participant responses. Drawing on speculative design techniques~\cite{auger2013speculative, dunne2024speculative, ouzounian2017speculative}, we asked participants to complete a three part user study process.
In particular, for each of the 10 pairs, we conducted two semi-structured interviews, in pairs, with the use of a design probe~\cite{wallace2013making, tahirouglu2020digital_musicinstruments_as_design_probes, fisher2016adapting_design_probes} after the first interview. 

The three parts are detailed as follows:

\begin{enumerate}

\item Interview 1: The first interview primarily discussed their current workflow using DAWs;

\item System usage: The second component included the use of our design probe in pairs (see Section C);

\item Interview 2: After using the design probe, we discussed real time music collaboration and the use of DAWs in MR. During this interview we asked them to speculate on what features and interaction techniques would be necessary to incorporate into a future MR-DAW. Specifically, what would be major challenges, necessary interactions, and features for such a system.

\end{enumerate}

We recorded, transcribed the interviews from the 10 pairs of participants, and coded them using inductive thematic analysis, rooted in grounded theory \cite{badreddin2013thematic,Hopkins2023Virtual}. Authors 1 and 2 independently coded and conducted the analysis of all 20 participants, which enabled us to create themes organically. We then discussed the topics considerably and converged on the most present topics in the dataset \cite{terry2017thematic,hopkins7brain}. MAXQDA \cite{kuckartz2019analyzing_MAXQDA} software was used to keep track of independent coding databases which facilitated discussion. Databases were merged after conversation to develop final themes. These themes are outlined in the following sections. 


\subsection*{Participants}

The study involved 20 participants with diverse musical backgrounds and professional experience in music creation and recording.

\textbf{Demographics:} The cohort was geographically distributed, with 12 participants from Taiwan and 8 from the United States. 12 participants identified as female and 8 as male. The age of participants ranged from 18 to 54 years (M = 30.22, SD = 7.85). The largest group was the 25--34 age bracket (n=16), followed by the 18--24 (n=2), 35--44 (n=1), and 45--54 (n=1) brackets.

\textbf{Musical Experience:} Participants had extensive musical experience, with a mean of 20.33 years (Median = 23.00). Their instrumental skills were varied. Piano and guitar were the most frequently cited instruments, often played alongside others such as violin, erhu, bass, percussion, and woodwinds (e.g., saxophone, flute, clarinet).

\textbf{Technology Proficiency:} Participants reported proficiency with a wide range of music technologies. Common DAWs included Logic Pro, Ableton Live, GarageBand, and FLStudio. Some participants also had experience with creative coding and sound design environments like Max/MSP and ChucK, as well as web-based music creation tools such as Suno and Chrome Music Lab. General experience in recording, editing, and working with interactive musical interfaces was common across the group.


\subsection{Interview 1: Current Use of Tools}
We began by asking our participants about their current workflow with respect to music composition and practice. This enabled us to get a better sense of the tools used by our musician participants to record, practice, and share music with each other. We extended our questioning to use of DAWs and collaborative co-creative activities in their musical practice. This was designed to engage the participants in reflection on their own practices and how technology is used to foster collaboration. Additionally, we discussed what tools they use or would use to collaborate on composition given the current state of available technologies. 

The emergent themes for this section are outlined below in no particular order of importance:

\subsubsection{\textbf{Asynchronous File Sharing}}
18 of 20 participants mentioned asynchronous file sharing as their current strategy for accomplishing collaborative songwriting when they are not co-located with their composition partners. With participants like P5 and P6 saying: ``Usually, I record my track, send it to my partner, and they add theirs later. It’s not real-time—we rely on post-editing." Additionally, participants indicated that they had a difficult time conceptualizing how they would do it any differently, given the current state of technology. Half of the participants also indicated explicitly that it would be great to have a ``Google docs for music editing", but they were unaware of any such technology.  

\subsubsection{\textbf{Collaboration Through Video Call}}
Half of the participants indicated that they would use a video conferencing application to accomplish co-creative song writing. With P19 suggesting ``there is video calling tools, which has not as good latency capabilities. But, you know, I'm sure we'd get something done." It was also observed that latency was a major concern for participants, with all participants mentioning the importance of low-latency in their remote musical collaborations, while only 1 (P20) mentioned ever personally using a low-latency service to collaborate musically. P19, P7, and P8, however, mentioned that in some cases they felt they needed time to come up with ideas on their own without sharing, and in some cases near-real-time music collaboration is undesirable, with P7 stating ``only merge the best parts."

\subsubsection{\textbf{Phone Recordings and Priority for Convenience}}
8 of 20 mentioned using the phone as their primary device for recording music in the initial stages of the recording process. These phone recordings were shared among band mates, or kept for personal recollection of musical ideas. Many mentioned the convenience being one of the number one factors. For example P3 stating: ``computers are bulky, phones are convenient." When it comes to capturing the musical moment, convenience and speed were of utmost importance, with P15 even creating their own term to describe priorities in their musical equipment: ``it should be couchable."  

\subsubsection{\textbf{Use of Large Screens for Recording, Mixing, and Mastering}}

All participants indicated that a computer or tablet is required to edit recordings beyond the draft stage. In contrast to the mobile screen, which suffices for drafting or capturing nascent musical ideas. P9, a professional composer in Taiwan, indicated that the songwriting company they work for requires the use of two computers and several screens in addition to an iPad. P9 described needing at least two screens to compose: ``I need two screens to work more efficiently." The tablet acts as an extension of the computer, showing virtual knobs and sliders. While the second computer runs on a separate operating system to offer dedicated resources for sound libraries. 

\subsubsection{\textbf{Constructing Workflow Based on Computer Placement}}
17 of 20 participants indicated that they place their computational device directly in front of them while recording. Suggesting that the computer, tablet, or mobile devices are central to the process of recording. Those that primarily use a computer suggested that they need to adapt their positioning to the use of a mouse and keyboard. With P20 explicitly stating: ``I guess instruments are better on the left...I can use my right hand to click around with a mouse at the same time." They indicated that the position is uncomfortable, but optimal given the technological requirements. Those that use a tablet or phone suggested that they need to use a dedicated music stand, table, or other surface to hold their device. These additional surfaces then become part of the workflow and are essential for the process of efficiently recording music. 

\subsubsection{\textbf{Use of Traditional Music Notation}}
Half of the participants also indicated relying heavily on traditional music notation. Whether it was writing music in traditional notation and transferring certain file types to a DAW later, or practicing and writing almost exclusively with traditional notation. 5 participants did not use software at all for the initial stages of writing and relied solely on the use of handwritten notation to formulate ideas. This was indicated as being done for speed and drafting. For example, P11 stating ``I would probably write it down first, and then after it's more complete, I would record it." Several also stating that they fear that they would forget the ideas if they don't write them down quickly enough, ``I would forget. I need to write it to be able to arrange it later."

\subsection{Reflections on the MR-DAW Design Probe}
During this portion of the study participants collaborated in pairs using our design probe in MR. Participants utilized a pedal to make recordings by pressing a “start recording” and “stop recording” button. After pressing the “stop recording” button the recorded section would play back in loops, mimicking the functionality of a loop pedal commonly used by musicians to construct solo compositions on-the-fly. Both participants had the ability to make recordings and contribute to the collective composition.

As recordings were being made, tracks were indicated as being present and could be stopped and started at will (see Figure \ref{fig:system_overview}C). This enabled manipulations of the composition by both participants simultaneously, affording the feeling of a co-creative composition process that was recorded and displayed using hand gestures and foot pedal presses.

The user interface panel that displayed tracks could be resized and moved around the environment, affording the spatial freedom to customize and rearrange their personal DAW configuration while affecting the same relative virtual objects. When interacting with virtual objects using the hands and feet a local version of Ableton Live 12 was being affected using Ableton’s builtin API. Thus, both participants were collaborating on a composition in Ableton that was being rendered and interacted with in MR.

Participants were located in separate rooms to reflect a networked composition process. Though directly wired to minimize latency and complexity of the system for testing, these processes are conceivable given near-real-time audio capabilities of current networked audio systems up to a certain distance (near 1000km).

\subsubsection{\textbf{Comparing Usability of Design Probe and Current Recording Technology}}

After participants utilized the MR-DAW system we asked them to complete a short NASA-TLX questionnaire on usability. This was aimed at understanding how the usability of the MRDAW system compared to their current workflow and use of technology. This helped us glean an understanding of how the participant perceived the capabilities of the MR-DAW and if it provided any benefit for usability (see Figure \ref{fig:nasa_tlx_comparison}).  

\section{NASA-TLX Workload Evaluation}

\begin{figure}[h]
  \centering
  \includegraphics[width=\linewidth]{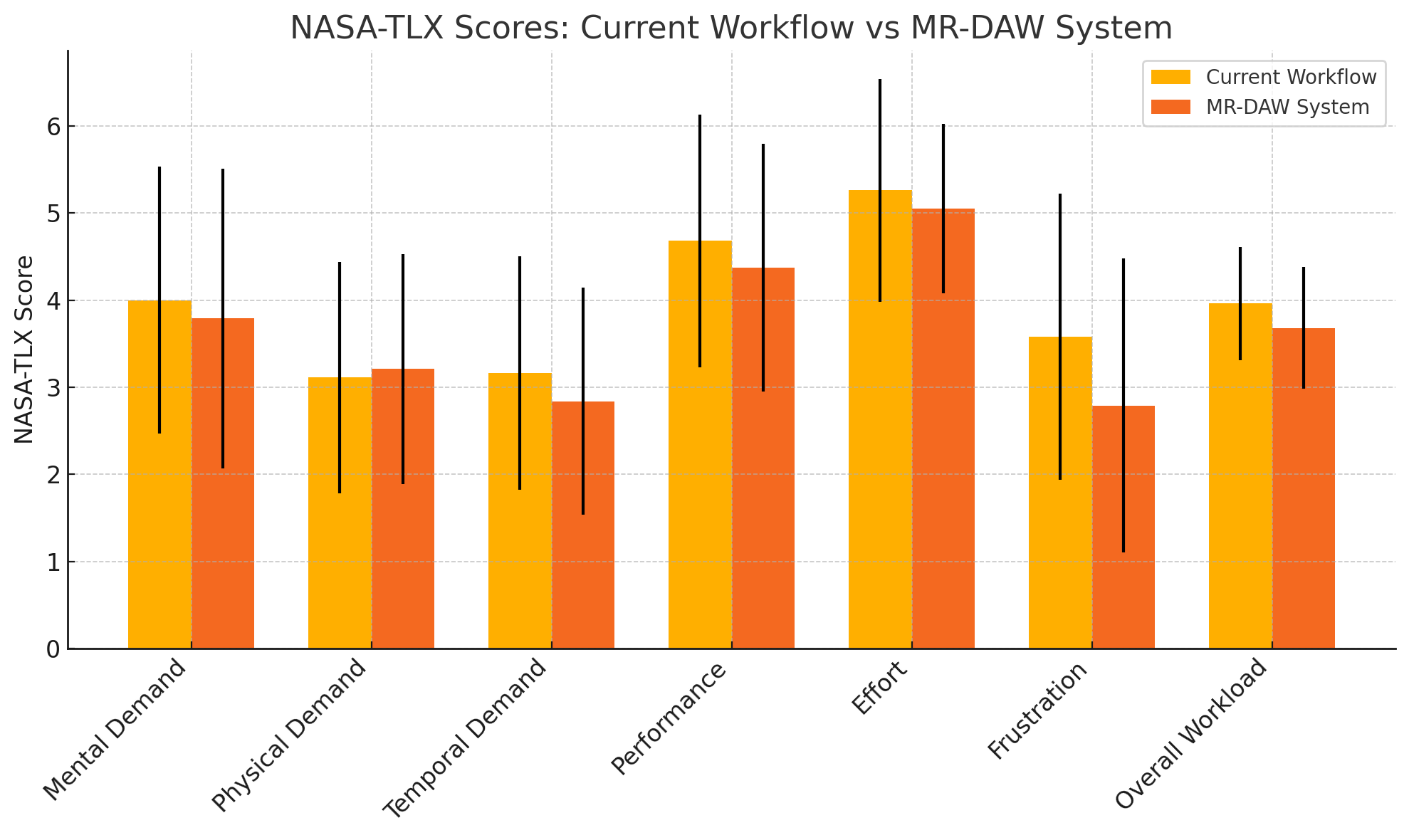}
  \caption{Comparison of NASA-TLX subscale scores between the Current Workflow and the MR-DAW System. Each bar represents the mean score for a subscale, with error bars indicating standard deviation. Lower scores indicate reduced workload.}
  \label{fig:nasa_tlx_comparison}
\end{figure}

To evaluate the perceived workload associated with the VRDAW system, we administered a NASA-TLX questionnaire to participants after they used both the VRDAW system and their current workflow. This analysis aimed to quantify the usability of the VRDAW system in comparison to existing technology and practices. The results of this analysis are presented in Figure~\ref{fig:nasa_tlx_comparison}.


The results show a trend towards a reduced cognitive workload when using the MR-DAW system. The mean overall workload score for the MR-DAW system was lower ($M=3.68, SD=0.70$) than that of the current workflow ($M=3.96, SD=0.65$). While this difference was not statistically significant, $t(18) = 1.74, p = 0.10$, the consistent pattern of lower scores suggests a positive impact on usability.

Examining the individual subscales, the most notable (though not statistically significant) decreases in workload were seen in the areas of \textit{Frustration} (a reduction from a mean of 3.58 to 2.79) and \textit{Temporal Demand} (a reduction from 3.16 to 2.84). This suggests that participants may have perceived the MR-DAW system as less irritating and the pace of the task as less hurried. Overall, while not conclusive, these findings indicate that the MR-DAW system has the potential to provide a tangible usability benefit by reducing the perceived effort and frustration involved in the task.


\subsection{Speculative Future of MR-DAWs}
The second interview asked participants to discuss their experiences using the system. We discussed what system components caused issues, were enjoyable, and would be desirable for future systems. Afterwards, we began a discussion speculating on the design of a future MR-DAW. We asked participants what the necessary challenges to overcome would be, components that should be retained or expanded, and what features would be desirable in future systems. We also discussed the possibilities for future systems and their use cases, outlining what new interactions would afford and how they might change their current workflow.

The emergent themes for this section are outlined below:

\subsubsection{\textbf{Spatial Control Enhancement in MR}}
16 of 20 participants explicitly mentioned the enhanced spatial freedom that an MR-DAW affords as compared to a traditional DAW. Having mentioned the freedom it would bring to be able to arrange their studio in exactly the way that they wanted and reconfigure at any point. P20 mentioned that this would change the way they operated in the studio because they could use MR ``in a way that you could never do on a computer, because you only have one input and output set."; and P15 stating: ``I always feel super claustrophobic when I'm working at home and it's like, I got the laptop there and  I'm never comfortable. And I feel like the power of VR is it's an entire computer on your head." 

P19 and P20, gigging and recording musicians, mentioned the ability to save on space in their recording studio by offloading the majority of their hardware to virtual objects and panels: ``in our garage we'd run out of space real quick, right? Because I got the computer over here. I have a massive audio interface over here, and then, a load of cable management that's going into that. It would become chaotic, very fast. The ability to take away a couple of components of that and just have them in VR. (It would add) mobility, modularity, and customizability." They both went on to discuss the benefits of collaboration in such a space: ``If you have a laundry list of things, and you trust the person that you're working with. (You could) take care of this half of the stuff in a way that you could never do on a computer, because you only have one like input and output set."

P19 also mentioned that they can adjust the angle of their screens in MR, which as discussed in the initial interview findings often requires an additional dedicated piece of hardware, stating: ``you could get that (virtual UI panels) into positions that you could never feasibly get a screen or a physical piece of equipment into, but that were really really convenient for recording."

Additionally, participants mentioned being able to better arrange tracks and effects in MR as compared to a traditional DAW: ``I can have all of the tracks that I'm working with, like, right over here. I have the effects bus over here. I have like the overall mixer over here. And, by the way, if I'm working with a specific track I can literally hold a given track into a different space." 

\subsubsection{\textbf{Alternative Use of Sensors for Control}}
All of our participants suggested that the use of MR creates the possibility for new methods of controlling the DAW. These included suggestions for head motion tracking, eye tracking, brain-computer interfaces, hand tracking, foot controllers, and voice commands. Some of these control mechanisms could be accomplished without the use of MR; however, participants mentioned that MR affords more flexibility than other devices for alternative controllers given less spatial constraints. For example, because a virtual content can be rendered in all directions several participants mentioned launching tracks like a conductor, with P20 stating: ``you could like raise the volume of like certain parts of the song, like as if you had like an orchestra in front of you."; and P20 suggesting: ``Could you imagine you organize all of the different tracks in space? And that's how you end up with dynamics. I guess if you're in the same VR space there's no reason why you couldn't direct and conduct even people from miles away."

\subsubsection{\textbf{Particularities of Each Instrument for Interaction}}
Our participants play a wide variety of instruments and we quickly discovered that some instruments are much more difficult to interact with in MR than others. For example, those who played violin had to hold their instrument using their neck with the additional weight of the VR headset, making it easy to fatigue. All guitar players also mentioned that they had a hard time seeing the neck of the guitar which negatively impacted their playing. 

This comes in contrast to vocalists which have full flexibility to use their hands to interact with virtual content. Additionally, for saxophonists, who do not rely on visual input, were not perturbed by the visual distortions caused by the VR headset. 

Similarly, the use of the foot pedal had unique hindrances for some instruments. For example, while playing cello, which is usually stabilized by the legs or knees of the player while bowing, the use a foot pedal creates instability of the instrument when the leg is lifted to press the pedal. However, it was suggested by one participant that the footpedal could be customized and controlled by the player. For example, in the case of looping, 10 of 20 participants suggested affording the ability to vocally initiate a count in for the initial loop. This would free up the functions made available on the foot pedal, which one participant suggested could be reassigned and customized in real time using a UI panel in MR. 

\subsubsection{\textbf{Considerations for Audio Routing}}
Six of the 20 participants explicitly mentioned the need for separate audio channels for communication and music. P17 and P18 suggested toggleable audio channels while recording in collaboration with another player so they could talk and hear the recording separately, suggesting: ``like a breakout room is the analogy that I'm kind of looking for. They don't leave, but audio wise, they hear just what they want to, like soloing."

\subsubsection{\textbf{Desire to See Face Rendering in MR}}
Similarly to audio routing considerations, 8 of the 20 participants mentioned the desire to see their collaborator's facial expressions. Though this is an ongoing challenge in MR generally, it underscores the importance of facial expressions and visual renderings in MR. Several participants mentioned that MR could easily afford this capability by enabling a screen that showed the other person, like a video call in MR. With P17 going as far as saying that without it there is a feeling of isolation in the space: ``does it kind of feel like you're just in this alone?"

\subsubsection{\textbf{Integration of Music Notation and Idea Capture}}
Considering the number of participants that indicated that they regularly utilize traditional music notation, it came as no surprise that many suggested that this was included in the DAW (8 of 20). Traditional DAWs often do not include traditional notation software, rather notation software is generally standalone. This requires musicians that utilize notation software to use both softwares concurrently and continuously trade files between programs. With P17 describing the process as ``clunky". 6 of 20 suggested automatic transcription software for MR, which is also an ongoing technical challenge in music technology, but the process of inputing traditional music notation using an MR device is also not a straight forward solution and could benefit greatly from automatic transcription. 

In addition to notation software, capturing musical ideas in the form of recorded video or audio was mentioned by several participants to capture initial creative moments. As described in the first interview, workshopping musical ideas is of great importance to the musicians, and doing so in MR would afford the ability to capture these ideas in myriad ways.

\subsubsection{\textbf{Creativity Enhancement}}
6 participants mentioned that our system ignites creativity primarily through collaborative interaction and the introduction of novel, external perspectives. Users find that the initial act of creating with others can lead to unexpected and interesting musical directions. P11 noted, ``The chord progression I played at the beginning was completely different from what I had originally imagined it would sound like. It's interesting to hear how others perceive the same chord progression and what they add to it." This highlights how the system facilitates a dynamic exchange of ideas, where one person's contribution serves as a catalyst for another's inspiration. This reciprocal process helps to overcome individual creative limitations, as  participant explained, "When you hear what others do, you suddenly get ideas about what you can add. It's like sparking creativity between two people." The immediate feedback inherent in the system is a powerful tool for unlocking novel creative pathways, with P5 stating, ``Maybe you didn't have that creativity initially, but after hearing someone else's immediate feedback, you get new creative ideas." Furthermore, even simple visual cues within the system can act as a potent starting point, as another user mentioned, ``That tree element we saw earlier really got us going." Ultimately, this collaborative environment not only sparks new ideas but also fosters a significant sense of achievement.

%% file: 05_Discussion.tex

The interviews yielded fruitful insights into how musicians use their equipment in a variety of musical practices, as well as what musicians may need from the future of music technology. We received an unexpected amount of excitement about the future of DAWs in MR, with all participants mentioning that MR would streamline practices in many ways. The modern music studio is almost always pictured with a giant screen in front of the engineer, and in the case of professional music composition and arrangement studios many practitioners use a myriad of screens and interactive devices in their workflows. 

Based on our findings, lay musicians seem to favor convenience and have spatial constraints. For example, many professional musicians have a recording studio in their homes; however, these studios take up an incredible amount of space--often a room or even floor--and are not feasible for lay musicians. Thus, a computer or tablet is favored, which imposes its own constraints in terms of interactions (screen size, finger-driven interactions, single input and output mechanisms, etc.). A future MR-DAW would solve these issues in an elegant manner, by providing lay musicians with a spatially unconstrained, portable, and customizable music studio. 

Furthermore, collaboration can be greatly enhanced with the use of MR for collaborative composition with the inclusion of several mentioned feature changes. For one, special consideration needs given to audio routing. It was brought to our attention that audio can get messy if one is trying to listen carefully while a recording is also playing. We foresee this necessitating further study to better understand how to communicate while playing and recording with another person. Also, being able to see your collaborator would enhance communication between the two persons. We also know from previous research \cite{hopkins2022late,HowLate_paper_ISMAR_9995407 ,ARDC_hopkins} that networked visual information has a greater tolerance for latency and the addition of video probably will not cause issues for collaborating musically.  

Nevertheless, it is worth noticing that to implement the vision of a low-latency collaborative DAW in MR, some critical technical issues need to be solved as recently highlighted in~\cite{boem2025issues}. These include the effective integration of NMP systems within Metaverse systems, or the advancement of hardware and software technologies to acquire, process and deliver music signals (including audio and MIDI) more efficiently.

We therefore call upon researchers focusing on the Musical Metaverse to consider testing and building components of an MR-DAW, which is bound to have many unforeseen hurdles to overcome. For example, many musicians described controlling musical elements with alternative controllers. We see this as a rich space for exploration, as assuming a standing position and using MR while playing frees the use of the hands, feet, adds sensors to the head, eye tracking, and front-facing cameras inherently. All of these and more are bound to open new doors for controlling musical elements for the purposes of recording and sharing new music with others. 

Though our comparative usability results did not show significance between the design probe (MR-DAW) and the users current workflow, there was a trend towards greater usability for the MR-DAW. We see this as users having time to adjust to and construct their own workflow, while we had them learn and use a non-optimized and new system for testing. The purpose of the probe was simply to demonstrate proof of concept and to spark thought and discussion between the participants. To this end, we think that an optimized and customizable MR-DAW interface in the future will be much more usable than current workflows.  

Additionally, we see many opportunities for MR technology to facilitate many musical activities. For example, when learning new music, musicians are often reading music notation or watching videos \cite{avila2019encumbered}, positioned at a desk, with there instruments in an awkward position in front of them. An MR-DAW would enable many musicians to assume a natural playing position and afford the freedom to place music notation or video feeds anywhere they would like while they continue their musical practice. Unfortunately, there are still many technical issues to overcome, such as visual distortion at certain angles of gaze, like the distorted guitar neck mentioned by participants. Also, one would need to consider how to render the virtual objects in a way that enables players to clearly see their hand with respect to virtual and real objects simultaneously--a serious issue for musicians raised by participants and specific to MR.


It is worth noticing that our study presents some limitations. Firstly, we involved only 10 pairs of musicians in our evaluation sessions. A greater number of participants could have improved the generalizability of our results.

Secondly, our user study was conducted in separate rooms to simulate a remote environment, but control messages to Ableton from both controllers were sent to the same machine via OSC over local host. Therefore, we did not test an infrastructure that handles control messages at a distance, limiting our ability to gauge how our system could deal with and how our participants would respond to even minimal delays. We see this as potentially the biggest hurdle to creating a near-real-time system given our approach. For our system to work well, control messages for initiating actions in Ableton need to be optimized. This function is beyond the scope of our current research as we felt providing a latency-free experience was the priority for getting our participants to be able to imagine such a speculative technology. 
Follow-up studies would further illuminate the capabilities and limitations of networked DAWs in MR. Therefore, we also suggest a study that explores a comparison in embodied components of such systems, such as incorporating other MIDI devices and controllers as a comparison to virtual elements. 


%% file: 06_Conclusion.tex
This paper investigates the potential of MR to overcome some barriers surrounding the use of DAWs, creating an intuitive environment for real-time, remote, musical collaboration. These barriers include primarily the fact that DAWs often encumber the musician's workflow, tethering them to a desk and hindering natural interaction with their instrument, and the fact that effective remote collaboration remains a significant challenge, with existing solutions hampered by network latency and asynchronous file sharing. 

Our study employed a speculative design approach to uncover the benefits and potential pitfalls of collaborative MR-DAWs for the Musical Metaverse. To facilitate discussion, we present MR-DAW, an MR system enabling multiple geographically dispersed users to control a single, shared DAW instance while moving freely in their local spaces. Such a networked system enables each remote musician to use a physical foot pedal for collaborative looping, merging a familiar, hands-free interaction with a shared virtual session. 

We evaluated the system through semi-structured interviews and user study involving 10 pairs of musicians. We analyzed current practices, reported on the experience of users interacting with our proposed MR system, and speculated on the future of musical collaboration in MR. Our results highlight the affordances of MR for unencumbered musical interaction and provide a speculative outlook on the future of remote collaborative DAWs in the Musical Metaverse.

%% file: IoMusT.bib
@inproceedings{abletonlink,
  title={Ableton Link--A technology to synchronize music software},
  author={Goltz, F.},
  booktitle={Linux Audio Conference},
  pages={39--42},
  year={2018}
}


%% file: NMP.bib
@article{rottondi2016overview,
	Author = {Rottondi, C. and Chafe, C. and Allocchio, C. and Sarti, A.},
	Journal = {IEEE Access},
	Publisher = {IEEE},
	Title = {An Overview on Networked Music Performance Technologies},
	Volume = {4},
	Pages = {8823--8843}, 
	Year = {2016}
}


%% file: Turchet.bib
@article{turchet2021MusicalXR,
  title={Music in Extended Realities},
  author={Turchet, L. and Hamilton, R. and {\c{C}}amci, A.},
  journal={IEEE Access},
  volume={9},
  pages={15810--15832},
  year={2021},
  publisher={IEEE}
}

@article{turchet2023musical,
  title={{Musical Metaverse: vision, opportunities, and challenges}},
  author={Turchet, Luca},
  journal={Personal and Ubiquitous Computing},
  pages={1--17},
  year={2023},
  publisher={Springer}
}

@article{boem2025issues,
  title={Issues and Challenges in Audio Technologies for the Musical Metaverse},
  author={Boem, Alberto and Tomasetti, Matteo and Turchet, Luca},
  journal={Journal of the Audio Engineering Society},
  volume={73}, 
  issue={3}, 
  pages={94--114},
  year={2025}
}

@article{turchet2023freesoundvr,
  title={FreesoundVR: soundscape composition in virtual reality using online sound repositories},
  author={Turchet, Luca and Carraro, Marco and Tomasetti, Matteo},
  journal={Virtual Reality},
  volume={27},
  number={2},
  pages={903--915},
  year={2023},
  publisher={Springer}
}


%% file: article_specific.bib
@inproceedings{avila2019encumbered,
  title={Encumbered interaction: a study of musicians preparing to perform},
  author={Avila, Juan Pablo Martinez and Greenhalgh, Chris and Hazzard, Adrian and Benford, Steve and Chamberlain, Alan},
  booktitle={Proceedings of the 2019 CHI conference on human factors in computing systems},
  pages={1--13},
  year={2019}
}

@inproceedings{barate2022collaborative,
  title={A collaborative digital audio workstation for young learners},
  author={Barate, Adriano and Ludovico, Luca A and Presti, Giorgio and others},
  booktitle={Proceedings of the 14th International Conference on Computer Supported Education},
  pages={458--464},
  year={2022},
}

@article{stickland2021design,
  title={Design and evaluation of a scalable real-time online digital audio workstation collaboration framework},
  author={Stickland, Scott and Athauda, Rukshan and Scott, Nathan},
  journal={Journal of the Audio Engineering Society},
  volume={69},
  number={6},
  pages={410--431},
  year={2021},
  publisher={Audio Engineering Society}
}

@article{marrington2017composing,
  title={Composing with the digital audio workstation},
  author={Marrington, Mark},
  journal={The singer-songwriter handbook},
  pages={77--89},
  year={2017},
  publisher={Bloomsbury New York}
}

@inproceedings{wallace2013making,
  title={Making design probes work},
  author={Wallace, Jayne and McCarthy, John and Wright, Peter C and Olivier, Patrick},
  booktitle={Proceedings of the SIGCHI conference on human factors in computing systems},
  pages={3441--3450},
  year={2013}
}

@article{auger2013speculative,
  title={Speculative design: crafting the speculation},
  author={Auger, James},
  journal={Digital Creativity},
  volume={24},
  number={1},
  pages={11--35},
  year={2013},
  publisher={Taylor \& Francis}
}


%% file: musical_XR.bib
@inproceedings{men2018lemo,
  title={LeMo: Supporting Collaborative Music Making in Virtual Reality},
  author={Men, L. and Bryan-Kinns, N.},
  booktitle={IEEE VR Workshop on Sonic Interactions for Virtual Environments},
  year={2018}
}

@inproceedings{cassorla2020augmented,
  title={{Augmented Reality for DAW-Based Spatial Audio Creation Using Smartphones}},
  author={Cassorla, A.M. and Kearney, G. and Hunt, A. and Riaz, H. and Stiles, M. and Murphy, D.T.},
  booktitle={Audio Engineering Society Convention 148},
  year={2020},
  organization={Audio Engineering Society}
}

@inproceedings{hopkins2022late,
  title={{How Late is Too Late? Effects of Network Latency on Audio-Visual Perception During AR Remote Musical Collaboration}},
  author={Hopkins, T. and Weng, S.C.C. and Vanukuru, R. and Wenzel, E. and Banic, A. and Do, E.Y.L.},
  booktitle={2022 IEEE Conference on Virtual Reality and 3D User Interfaces Abstracts and Workshops},
  pages={686--687},
  year={2022},
  organization={IEEE}
}


%% file: other.bib
@article{Hopkins2023Virtual,
	author = {Hopkins, Torin and Jude, Alvin and Phillips, Greg and Do, Ellen Yi-Luen},
	journal = {AIMC 2023},
	year = {2023},
	month = {aug 29},
	note = {https://aimc2023.pubpub.org/pub/9064kmvw},
	publisher = {},
	title = {Virtual {AI} {Jam}: AI-{Driven} {Virtual} {Musicians} for {Human}-in-the-{Loop} {Musical} {Improvisation}    },
}

@article{ARDC_hopkins,
AUTHOR={Hopkins, Torin and Weng, Suibi Che Chuan and Vanukuru, Rishi and Wenzel, Emma A. and Banic, Amy and Gross, Mark D. and Do, Ellen Yi-Luen},    
TITLE={{AR Drum Circle: Real-Time Collaborative Drumming in AR}},     
JOURNAL={Frontiers in Virtual Reality},      
VOLUME={3},           
YEAR={2022}   
}

@inproceedings{hopkins2024xrjam,
  title={XR Jam: Design Considerations for Music Networking with AI in XR},
  author={Hopkins, Torin and Weng, Suibi Che Chuan and Vanukuru, Rishi and Novack, Sasha and Tobin, Chad and Wenzel, Emma and Banic, Amy and Gross, Mark and Do, Ellen Yi-Luen},
  booktitle={2024 IEEE International Conference on Artificial Intelligence and eXtended and Virtual Reality (AIxVR)},
  pages={97--101},
  year={2024},
  organization={IEEE}
}

@inproceedings{jones2023abletonosc,
  title={AbletonOSC: A unified control API for Ableton Live},
  author={Jones, Daniel},
  booktitle={Proceedings of the International Conference on New Interfaces for Musical Expression, Miguel Ortiz and Adnan Marquez-Borbon (Eds.). Mexico City, Mexico, Article},
  volume={60},
  number={5},
  year={2023}
}

@book{dunne2024speculative,
  title={Speculative Everything, With a new preface by the authors: Design, Fiction, and Social Dreaming},
  author={Dunne, Anthony and Raby, Fiona},
  year={2024},
  publisher={MIT press}
}

@article{tahirouglu2020digital_musicinstruments_as_design_probes,
  title={Digital Musical Instruments as Probes: How computation changes the mode-of-being of musical instruments},
  author={Tah{\i}ro{\u{g}}lu, Koray and Magnusson, Thor and Parkinson, Adam and Garrelfs, Iris and Tanaka, Atau},
  journal={Organised Sound},
  volume={25},
  number={1},
  pages={64--74},
  year={2020},
  publisher={Cambridge University Press}
}

@article{badreddin2013thematic,
  title={Thematic review and analysis of grounded theory application in software engineering},
  author={Badreddin, Omar},
  journal={Advances in Software Engineering},
  volume={2013},
  number={1},
  pages={468021},
  year={2013},
  publisher={Wiley Online Library}
}

@article{terry2017thematic,
  title={Thematic analysis},
  author={Terry, Gareth and Hayfield, Nikki and Clarke, Victoria and Braun, Virginia and others},
  journal={The SAGE handbook of qualitative research in psychology},
  volume={2},
  number={17-37},
  pages={25},
  year={2017},
  publisher={SAGE Publications Ltd}
}

@article{hopkins7brain,
  title={BrAIn Jam: Neural Signal-Informed Adaptive System for Drumming Collaboration with an AI-Driven Virtual Musician},
  author={Hopkins, Torin and Sun, Ruojia and Weng, Suibi Che Chuan and Ma, Shih-Yu and Crum, James and Hirshfield, Leanne and Do, Ellen Yi-Luen},
  journal={Frontiers in Computer Science},
  volume={7},
  pages={1570249},
  publisher={Frontiers}
}

@INPROCEEDINGS{HowLate_paper_ISMAR_9995407,
  author={Hopkins, Torin and Weng, Suibi Che-Chuan and Vanukuru, Rishi and Wenzel, Emma and Banic, Amy and Gross, Mark D and Do, Ellen Yi-Luen},
  booktitle={2022 IEEE International Symposium on Mixed and Augmented Reality (ISMAR)}, 
  title={Studying the Effects of Network Latency on Audio-Visual Perception During an AR Musical Task}, 
  year={2022},
  volume={},
  number={},
  pages={26-34},
  doi={10.1109/ISMAR55827.2022.00016}}

@article{ouzounian2017speculative,
  title={Speculative designs: towards a social music},
  author={Ouzounian, Gascia and Haworth, Christopher and Bennett, Peter},
  year={2017},
  publisher={Michigan Publishing}
}

@inproceedings{fisher2016adapting_design_probes,
  title={Adapting design thinking and cultural probes to the experiences of immigrant youth: Uncovering the roles of visual media and music in ICT wayfaring},
  author={Fisher, Karen E and Yefimova, Katya and Bishop, Ann P},
  booktitle={Proceedings of the 2016 CHI conference extended abstracts on human factors in computing systems},
  pages={859--871},
  year={2016}
}

@book{kuckartz2019analyzing_MAXQDA,
  title={Analyzing qualitative data with MAXQDA},
  author={Kuckartz, Udo and R{\"a}diker, Stefan},
  year={2019},
  publisher={Springer}
}
